\documentclass[10pt, conference, letterpaper]{IEEEtran}
\usepackage{cite}
\usepackage{hyperref}
\usepackage{amsmath}
\usepackage{amssymb}
\usepackage{optidef}
\usepackage{physics}
\usepackage{amsthm}
\usepackage{algorithm}
\usepackage{algpseudocode}
\usepackage{booktabs}
\usepackage{tabularx}
\usepackage{caption}
\usepackage{subfigure}

\algrenewcommand\algorithmicrequire{\textbf{Input:}}
\algrenewcommand\algorithmicensure{\textbf{Output:}}

\begin{document}

\title{Multi-Criteria Client Selection and Scheduling with Fairness Guarantee for Federated Learning Service}
\author{
\IEEEauthorblockN{Meiying Zhang\IEEEauthorrefmark{1}, Huan Zhao\IEEEauthorrefmark{1}, Sheldon Ebron\IEEEauthorrefmark{1}, Ruitao Xie\IEEEauthorrefmark{2}, Kan Yang\IEEEauthorrefmark{1}}

\IEEEauthorblockA{\IEEEauthorrefmark{1}Dept. of Computer Science, University of Memphis,  USA. 
\{mzhang6, hzhao2, sebron, kan.yang\}@memphis.edu}
\IEEEauthorblockA{\IEEEauthorrefmark{2}College of Computer Science and Software Engineering, Shenzhen University, China. drtxie@gmail.com}
}

\maketitle

\begin{abstract}
Federated Learning (FL) enables multiple clients to train machine learning models collaboratively without sharing the raw training data. However, for a given FL task, how to select a group of appropriate clients fairly becomes a challenging problem due to budget restrictions and client heterogeneity. In this paper, we propose a multi-criteria client selection and scheduling scheme with a fairness guarantee, comprising two stages: 1) preliminary client pool selection, and 2) per-round client scheduling. Specifically, we first define a client selection metric informed by several criteria, such as client resources, data quality, and client behaviors. Then, we formulate the initial client pool selection problem into an optimization problem that aims to maximize the overall scores of selected clients within a given budget and propose a greedy algorithm to solve it. To guarantee fairness, we further formulate the per-round client scheduling problem and propose a heuristic algorithm to divide the client pool into several subsets such that every client is selected at least once while guaranteeing that the `integrated' dataset in a subset is close to an independent and identical distribution (iid). Our experimental results show that our scheme can improve the model quality especially when data are non-iid.
\end{abstract}

\begin{IEEEkeywords}
    Federated Learning, Client Selection, Client Scheduling, Fairness, Behavior, Resource, Data Quality
\end{IEEEkeywords}

\section{Introduction}
In traditional machine learning, the training data are usually hosted by a centralized server (e.g., cloud server) running the learning algorithm or shared to a set of participating nodes for distributed learning \cite{meng2016mllib}. 
However, in many applications, the data cannot be shared to the cloud or other participating nodes due to privacy or legal restrictions, especially when multiple organizations are involved. 
Federated Learning (FL) enables multiple parties to train machine learning models collaboratively without sharing the raw training data \cite{mcmahan2017communication, kairouz2019advances}. 
All parties train the model on their local datasets and send the local model updates to an aggregator, who will aggregate all the local model updates and send the global model to each client for the next round of training until converge. 

Due to the privacy-preserving nature and the decentralized structure, the FL framework can be widely applied in many AI-driven applications where data are sensitive or legally restricted, such as smart healthcare \cite{perveen2019prognostic, chekroud2016cross}, smart transportation \cite{sallab2017deep,  najjar2017combining}, smart finance \cite{fiore2019using, babaev2019rnn}, and smart life \cite{shi2017deep, hino2018machine}. However, when deploying FL in these applications, it is challenging to find a group of clients who have the corresponding datasets and are willing to participate in an FL task. 


To cope with this challenge, we envision an FL service provider that will provide FL as a service for different applications (e.g., FLaaS\cite{kourtellis2020flaas}). In an FL service system, FL task requesters (i.e., customers) send different types of FL tasks to the FL service provider with some requirements (e.g., datasets and budgets). The FL service provider will recruit an appropriate group of clients who can satisfy the task requirements. Due to the heterogeneity of clients (including computing and communication resources, dataset size, and data quality), different clients may ask for different per-round prices for an FL task. Moreover, even if a client is selected for an FL task, it does not mean the client is able to participate in all the FL rounds because a) only a subset of clients will be selected from the client pool to participate in an individual FL round in order to reduce the communication and computation costs and b) a client may be unavailable during several FL rounds due to conflicting scheduling, out-of-battery, unstable networking environments. 
This makes the client selection a challenging problem in the FL service. 
Specifically, a promising client selection solution should consider both \textit{initial client pool selection} and \textit{client selection in each FL round}
with the following requirements: 

\begin{itemize}
    \item 
\textbf{Model Quality} The main goal of the client selection is to maximize the quality of the final model by selecting well-performed clients within the total budget. However, the actual performance is unknown during the initial client pool selection for an FL task. Moreover, many existing client selection algorithms \cite{nishio2019client, abdulrahman2020fedmccs, wang2020optimizing, li2020federated, fraboni2021clustered} only focus on the per-round client selection for a given FL task.

\item \textbf{Fairness} The fairness consists of two parts: 1) each client who satisfied the FL task requirement has the chance to be selected into the initial client pool for an FL task; and 2) each client has a similar chance to be selected during a given FL task. 
Many existing client selection methods may prefer a specific type of clients (e.g., with high computing/communication resources \cite{nishio2019client, abdulrahman2020fedmccs}, low non-iid data distribution \cite{li2022data, wolfrath2022haccs}, or high model performance \cite{chen2020optimal}), which makes it unfair for those clients whose resource or data quality may not be the best, especially for FL services. In \cite{huang2020efficiency}, the authors consider fairness during the client selection and aim to minimize the average model exchange time subjecting to long-term fairness guarantee and client availability.
\end{itemize}

In this paper, we propose a multi-criteria client selection and scheduling scheme with a fairness guarantee for FL service which consists of two stages: 1) selecting an initial client pool for each FL task by mapping requirements from the FL requester with the client capabilities (e.g., resources, data, previous behaviors, prices) under a given budget; and 2) selecting a subgroup of clients in each learning round of an FL task based on the data quality, model quality and behavior in previous rounds of this FL task. Specifically, the client pool will be divided into several subsets, and each subset will take turns participating in one training round. We refer to this set of rounds as one scheduling period in which all subsets have participated once. Scheduling periods are repeated until the global model converges. In each scheduling period, we compute a reputation score for each participating client according to their performance measured by model quality and behavior. Clients who are unavailable during the next scheduling period or have low reputation scores in the previous scheduling period are temporarily removed from the client pool and added back after one or a few scheduling periods. 

The contributions of this paper are summarized as follows:
\begin{itemize}
    \item We define a client selection metric based on multiple criteria, including 1) client resources (e.g., CPU, GPU, memory, power, communication bandwidth, time availability); 2) data quality (e.g., dataset size, data distribution); and 3) client behaviors (e.g., number of FL tasks completed, model performance in previous FL tasks, and dropoff ratio, etc.). We further define a score function to quantify each criterion. 
    
    \item We formulate the initial client pool selection problem into an optimization problem that aims to maximize the overall scores of selected clients in the initial client pool within a given budget from the task requester, subjecting to a minimal number of clients constraint and a minimum requirement for each of the criterion scores of every client. We also propose an efficient greedy algorithm to solve this optimization problem.

    \item To guarantee fairness, we further formulate the  per-round client selection into a client scheduling problem. Then, we propose a heuristic algorithm to divide the client pool into several subsets such that each client can participate at least once, while guaranteeing that the `integrated' dataset in a subset is close to iid distribution. 
 
    \item We show that our proposed client scheduling algorithms can guarantee fairness. The experimental results of training CNN models on MNIST and CIFAR datasets demonstrate that our scheme can improve the model quality especially when data are non-iid.
\end{itemize}

 The remainder of this paper is organized as follows. In \ref{sec:relatedwork}, we present the related work of client selection in FL. Section \ref{sec:systemmodel} describes the system model of an FL service system and the basic training process of an FL task. A client selection metric is defined in Section \ref{sec:metric}, followed by the problem formulation in Section \ref{sec:formulation}. In Section \ref{sec:solution}, we propose a solution for each formulated problem. Section \ref{sec:fairness} provides the fairness analysis of our proposed client selection solution, and 
 Section \ref{sec:evaluation} provides experimental evaluation. Finally, Section \ref{sec:conclusion} concludes the paper and provides future work.




\section{Related Work}\label{sec:relatedwork}
In \cite{nishio2019client}, the authors formulated a client selection problem to maximize the number of selected clients based on computation and communication resources when applying FL in the mobile edge computing framework. A greedy algorithm is proposed to solve the client selection problem. 
In \cite{abdulrahman2020fedmccs}, a multicriteria-based client selection approach (FedMCCS) is proposed to maximize the number of clients selected in each FL round. FedMCCS first filters the clients based on time, then uses a linear regression model to predict whether a client is able to perform the FL task based on the CPU, memory, and energy. 

These methods \cite{nishio2019client, abdulrahman2020fedmccs} require an extra round of communication of the resource information from clients to the server. Moreover, maximizing the number of clients selected in each FL round will output a better result and also require a higher cost. However, in FL services, the budget from an FL task request is usually limited. In \cite{qu2022context}, a client's successful participation probability is estimated based on the context (e.g., CPU frequency, RAM, storage, and channel information) in the current and all the previous rounds. Then, an optimization problem is formulated to select the clients within a given budget for a hierarchical FL. 

Randomly selecting a subset of clients (e.g., FedAvg) or selecting clients based on their computing and communication resources may lead to biased results. To guarantee unbiasedness, Li \emph{et al.} proposed an unbiased sampling scheme based on a multinomial distribution (MD) where client probabilities correspond to their relative sample size \cite{li2020federated}.
Although MD sampling can guarantee unbiasedness in expectation, it may still cause high variance in the amount of times a client is selected in a single FL iteration.  
In \cite{fraboni2021clustered}, two clustered sampling algorithms (based on sample size and model similarity) are introduced to reduce variance and keep unbiasedness during the client selection in FL. 

Many existing client selection methods may prefer a specific type of client (e.g., with high computing/communication resources \cite{nishio2019client, abdulrahman2020fedmccs}, low non-iid data distribution \cite{li2022data, wang2020optimizing, zhang2021client}, or high model performance \cite{chen2020optimal, cho2022towards}), which makes it unfair for those clients whose resource or data quality may not be the best, especially for FL services. In \cite{huang2020efficiency}, the authors consider fairness during the client selection and aim to minimize the average model exchange time subjecting to long-term fairness guarantee and client availability. This optimization assumes that the maximal number of clients in each FL round is fixed.

\section{System Model}\label{sec:systemmodel}
In this section, we first describe the system model of the FL service system. Then, we describe the basic training process of an FL task. 

\begin{figure}[!t]
    \centering
    \includegraphics[width=0.48\textwidth]{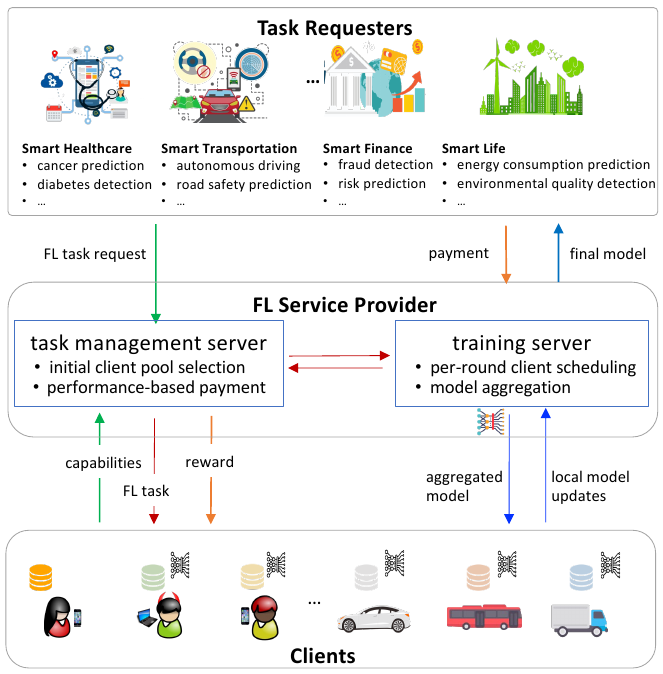}
    \caption{System Model of FL Services}
    \label{fig:systemmodel}
\end{figure}

\textbf{FL Service}: As shown in Fig. \ref{fig:systemmodel}, an FL service system consists of the following entities: 
\begin{itemize}
    \item \textit{Task Requester}: Task requesters (i.e., customers) send different types of FL tasks to the FL service provider with some requirements (e.g., datasets, models, minimum computing/communication requirement, time, budget, etc.).

    \item \textit{Clients}: Clients register with the FL service provider to participate in some FL tasks for profit. Each client will ask for a per-round price (the price may vary for different types of FL tasks) and the capabilities in terms of computing/communication resources, available time, and datasets. 
    
    \item \textit{Service Provider}: The FL service provider will recruit an appropriate group of clients who can satisfy the task requirements. However, even if a client is selected for an FL task, it does not mean the client is able to participate in all the FL rounds because a) only a subset of clients will be selected from the client pool to participate in an individual FL round in order to reduce the communication and computation costs and b) a client may be unavailable during several FL rounds due to conflicting scheduling, out-of-battery, unstable networking environments.
\end{itemize}

\textbf{FL Training Process}: 
A general FL training process happens between an aggregator and a set of clients $S$. Let $\mathcal{D}_k$ be the local dataset held by the client $k~(k\in S)$. The typical FL goal \cite{mcmahan2017communication} is to learn a model collaboratively without sharing local datasets by solving 
\begin{equation}
    \min_{w} F(w)  = \sum_{k\in S} p_k\cdot F_k(w),~s.t.~\sum_{k\in S} p_k  = 1 ~(p_k\geq 0),
\end{equation}
where $F_k(w) = \frac{1}{n_k}\sum_{j_k = 1}^{n_k} f_{j_k}(w; x^{(j_k)}, y^{(j_k)})$ is the local objective function for client $k$ with $n_k = |\mathcal{D}_k|$ available samples. $p_k$ is usually set as $p_k = n_k/\sum_{k\in S} n_k$ (e.g., FedAvg \cite{mcmahan2017communication}). 
An FL training process usually contains multiple rounds, and a typical FL round consists of the following steps: 
\begin{enumerate}
    \item \textit{client selection and model update}: a subset of clients $S_t$ is selected, each of which retrieves the current global model $w_t$ from the aggregator.
    \item \textit{local training}: each client $k$ trains an updated model $w^{(k)}_t$ with the local dataset $\mathcal{D}_k$ and shares the model update $\Delta_t^{(k)} = w_t - w_t^{(k)}$ to the aggregator.
    \item \textit{model aggregation}: the aggregator computes the global model updates as $\Delta_t = \sum_{k\in S_t} p_k \Delta_t^{(k)}$ and update the global model as $w_{t+1} = w_t - \eta \Delta_t$, where $\eta$ is the server learning rate. 
\end{enumerate}

\section{Client Selection  Criteria}\label{sec:metric}
As aforementioned, clients can be chosen based on various selection criteria including measures of computation resources (e.g., CPU, memory, storage, and battery/energy), communication resources (e.g., connection status, network bandwidth), data quality (e.g., dataset size and distribution) and reputation measures (e.g., historical model quality and behavior). 
To assist client selection in FL services, we first define a comprehensive client selection metric that consists of all of these factors. Specifically, 
 we define a score for each of these criteria and represent them as a vector \( \vectorbold*{s}=(s_{criterion,1},s_{criterion,2},...,s_{criterion,\#}) \). 
where \# is the number of criteria. Each score \( s_{criterion,i} \) is scaled to (0, 1). 

\begin{table}[!h]
    \centering
    \caption{Notations}
    \begin{tabular}{|c|c|c|} 
     \hline
     \textbf{Score} & \textbf{Description} \\ 
     \hline
     \(s_{CPU}\) & Available CPU ratio \\ 
     \hline
     \(s_{GPU}\) & Available GPU ratio \\ 
     \hline
     \(s_{MEM}\) & Memory size \\ 
     \hline
     \(s_{STR}\) & Storage size \\ 
     \hline
     \(s_{POW}\) & Power of the device (e.g., battery life) \\ 
     \hline
     \(s_{BDW}\) & Network bandwidth \\ 
     \hline
     \(s_{CON}\) & Connection status \\ 
     \hline
     \(s_{DataSize}\) & Data size \\ 
     \hline
     \(s_{DataDist}\) & Data distribution \\ 
     \hline
     \(s_{ModelQ}\) & Historical model quality \\ 
     \hline
     \(s_{Bhvr}\) & Behavior score \\ 
     \hline
    \end{tabular}
    \label{table1}
\end{table}

 Table \ref{table1} summarizes the criteria used in this paper.  
Depending on the specific FL task, the server can choose one or more of the listed scores or add scores for additional criteria to consider in client selection. Optionally, the server can also combine two or more of the scores into one score as necessary for ease of computation or simplicity. For example, scores for CPU, GPU, MEM, storage, and power can be combined into a single score for computation resources by applying weighted sum. 


\subsection{Definition of Resource Scores}
The computation and communication resources can be obtained during the client registration or when submitting client preferences for the FL tasks. 
We can convert the computation and communication information into the scores by comparing them with the minimal requirements of an FL task. 
Suppose for a given FL task, the minimal requirement of the computation and communication resources are defined by the FL task requester as 
$(CPU_{min}, GPU_{min}, MEM_{min}, Storage_{min}, POW_{min}$,
$BDW{min}, CON_{min})$, the score of CPU for client $i$ can be calculated as $ s_{CPU, i} = CPU_{client_i} / CPU_{min} $. 

Then, all the $s_{CPU, i} (i\in [1, n])$ are normalized into the range of (0, 1). Similarly, \(s_{GPU}\), \(s_{MEM}\), \(s_{STR}\), \(s_{POW}\), 
$s_{BDW}$, \(s_{CON}\) can be calculated by the FL service provider. We can use a similar approach to compute the score $s_{DataSize}$. 
 Other approaches can also be applied in calculating resource scores. For example, the connection channel status \(s_{CON}\) can also be measured using Shannon's equation as in \cite{qu2022context}. 

\subsection{Definition of Data Distribution Score $s_{DataDist}$}
The data distribution score \(s_{DataDist}\) characterizes how the client's data is independently and identically distributed (iid). It is defined as the complement of non-iid degree \(Nid\). That is, $s_{DataDist} = 1 - Nid$ 
where \(Nid\) is a function of a histogram \(\vectorbold*{h}=(h_1,h_2,...,h_c)\) which represents distribution of a client's data over classes \(1,2,...,c\) of a classification task. For example, \(\vectorbold*{h}=(10,20,...)\) means the client has 10 data samples of the first class label, 20 data samples of the second class label, and so on. We define the non-iid degree as the fraction of difference between sample sizes of maximum and minimum classes over the total sample size.
\begin{equation} \label{eq:nid}
    Nid(\vectorbold*{h}) = (max(\vectorbold*{h}) - min(\vectorbold*{h})) / sum(\vectorbold*{h})
\end{equation}

Alternatively, the non-iid degree can also be defined as a distance between two distributions, i.e., the client's data distribution \(\vectorbold*{h}\) and the uniform distribution \(\vectorbold*{u}=(\frac{1}{c},\frac{1}{c},...,\frac{1}{c})\), such as L2 distance \cite{li2022data}, Hellinger distance \cite{wolfrath2022haccs}, Kullback–Leibler (KL) divergence \cite{lee2022data}, etc..

\subsection{Definition of Historical Model Quality Score \(s_{ModelQ}\)}
The FL service provider will maintain historical model quality evaluations for each FL task and every client who participated in the task. That is, for each client, there is a vector \( \vectorbold*{q} = (q_{task_1}, q_{task_2},...) \), where \(q_{task_i}\) represents the quality of the model computed by this client for task \(i\). This per-task model quality \(q_{task_i}\)is the average of per-round model quality values \(q_t\) for all rounds in which the client successfully participated. Thus, for each client, 
\begin{equation} \label{q_task}
    q_{task_i} = \frac{1}{|\mathbb{T}|}\Sigma_{t \in \mathbb{T}} q_t
\end{equation}
where \(\mathbb{T}\) is the set of indices of participated rounds. 

For each round \(t\), the model quality value \(q_t\) can be defined as a cosine or other similarity between the local and global models: $ q_t = sim(\vectorbold*{w_l},\vectorbold*{w_g}) $ 
where \(\vectorbold*{w_l}\) is the local model parameter vector (model update) computed by the client, \(\vectorbold*{w_g}\) is the global model parameter vector obtained after the aggregation step for this round. 

The historical model quality score \(s_{ModelQ}\) is defined as the average of all \(q_{task_1}, q_{task_2},...\). The server can maintain model qualities for a fixed number $|\vectorbold*{q}|$ of recently completed tasks as $ s_{ModelQ} = (q_{task_1} + q_{task_2} + ...) / |\vectorbold*{q}| $.

\subsection{Definition of Behavior Score \(s_{Bhvr}\)}
The behavior score measures how often a client drops out or fails to return a model in an FL task. In the same method as the model quality score is defined, a per-round behavior score \(b_t\) is computed for each participating client after each round of a task. The per-round behavior score \(b_t\) is a binary indicator of whether the client successfully returns its local model update to the server:
\begin{equation} \label{b_t}
    b_t =
    \begin{cases}
      1 & \text{if model update is successful}\\
      0 & \text{else}
    \end{cases} 
\end{equation}

 The per-round behavior scores are averaged over all rounds in which the client has participated to obtain a per-task behavior score:
 \begin{equation} \label{b_task}
     b_{task_i} = \frac{1}{|\mathbb{T}|}\Sigma_{t \in \mathbb{T}} b_t.
 \end{equation}
 which is maintained in a vector \( \vectorbold*{b}=(b_{task_1},b_{task_2}) \). The overall behavior score \(s_{Bhvr}\) is the average of all or recent per-task behavior scores: $ s_{Bhvr} = (b_{task_1} + b_{task_2} + ...) / |\vectorbold*{b}| $

\subsection{Overall Score and Cost}
For convenience, we rewrite the scores in Table \ref{table1} \(s_{CPU}, s_{GPU}, ..., s_{Bhvr}\) as \(s_1, s_2, ..., s_{11}\). So, the score vector \(\vectorbold*{s}\) is written as 
   $ \vectorbold*{s} = (s_1, s_2, ..., s_{11}) $.
We take a weighted or unweighted sum of all the above scores as an overall score \(Score\) to denote how good a client is overall in all aspects:
\begin{equation} \label{overallscore}
        Score = Score(\vectorbold*{s}) = \vectorbold*{w} \cdot \vectorbold*{s}  = \sum_{i=1}^{11} w_i s_i 
\end{equation}
where the weight vector \(\vectorbold*{w}\) can be defined by the server according to the specific requirements of the FL task. 

Since clients contribute their resources and time when participating in an FL task, they also ask for a price (e.g., per-task price) for the FL task. Here we define the price as $Cost$. For example, the \(Cost\) can be determined by a function of the overall score of a client:
\begin{equation} \label{cost}
    Cost = Cost(Score) = aScore + b
\end{equation}
where \(a, b\) are constants, \(a>0\). 
Alternatively, \(Cost\) can be defined as any increasing function of the overall score \(Score\), or a combination of one or more of the separate scores \(s_1, s_2, ...\). Also, each client may give a price of its own as a constant.  In general, clients with higher scores demand higher costs. The maximal total cost for the selected clients is subject to a budget defined by the FL task requester. 


\section{Problem Formulation}\label{sec:formulation}
In this section, we formulate the client selection and scheduling problem based on the criteria defined in the previous section, which consists of two stages: 1) initial client pool selection, and 2) per-round client scheduling. 

\subsection{Stage 1: Initial Client Pool Selection for an FL Task}

Given an FL task, let \(\mathbb{K}\) denote the set of all clients willing to participate, and \(\mathbb{S} \subseteq{\mathbb{K}} \) the set of selected clients for this task. At Stage 1, we aim to build a client pool from which a subset is selected to participate in each round. We require that the number of selected clients is at least \(n^*\). The aim is to select high-score clients as many as possible within a limited cost budget \(B\). To do this, we want to maximize the sum of overall scores of selected clients with a budget constraint and a number-of-clients constraint. We also enforce a minimum requirement \(\vectorbold*{s_{th}}=(s_{1,th},s_{2,th},...,s_{9,th})\) for each of the criterion scores \(\vectorbold*{s_k}=(s_1,s_2,...,s_9)\) of each client \(k\). We can formulate the stage-1 client selection problem as follows. 
\begin{maxi!}|s|
{\mathbb{S}\subseteq{\mathbb{K}}}{\sum_{k\in \mathbb{S}}{Score_k}\label{obj1}}
{\label{P1}}{}
\addConstraint{\sum_{k\in \mathbb{S}}{Cost_k} \leq B \label{const11}}
\addConstraint{|\mathbb{S}| \geq n^* \label{const12}}
\addConstraint{\vectorbold*{s_k} \geq \vectorbold*{s_{th}}, \forall k \in \mathbb{S}\label{const13}}
\end{maxi!}

\subsection{Stage 2: Per-round Client Scheduling} \label{stage2_def}
At Stage 2, we will partition the client pool \(\mathbb{S}\) into \(T\) subsets and schedule a subset to participate in each training round such that the subsets take turns to participate in rounds \(1,2,..., T\). 
We refer to these \(T\) rounds as one scheduling period in which all subsets have participated once. Scheduling periods are repeated until the global model converges. In each scheduling period, we compute a reputation score for each participating client according to their performance measured by model quality and behavior. Clients who are unavailable during the next scheduling period or have low reputation scores in the previous scheduling period are temporarily removed from the client pool and added back after one or a few scheduling periods. More specifically, given an initial client pool $\mathbb{S}$ selected at Stage 1, each scheduling period includes the following steps: 
\begin{itemize}
    \item Step 1: Generate subsets \(\mathbb{S}_1, \mathbb{S}_2, ... , \mathbb{S}_T \subset \mathbb{S}\).
    \item Step 2: For \(t=1,2,...,T\), all the clients in the subset \(\mathbb{S}_t\) participate in round \(t\), update per-round model quality scores and per-round behavior scores for clients in \(\mathbb{S}_t\).
    \item Step 3: Update reputation scores and availability information for all clients in \(\mathbb{S}\).
    \item Step 4: Update client pool \(\mathbb{S}\) according to their  reputation scores or updated availability, including 
    \begin{itemize}
        \item removing clients that are unavailable in the next scheduling period; 
        \item removing clients that have bad reputation scores in the current period;
        \item adding clients that have been suspended for a fixed number of rounds due to bad performance.
    \end{itemize} 
\end{itemize}

The reputation score is defined as the sum of per-task model quality score and per-task behavior score as defined in Equations \ref{q_task} and \ref{b_task}: $s_{rep} = q_{task} + b_{task}$. 


\textbf{Subset Generation Problem:}
In the subset generation step, our goal is to let the union of all the subsets cover the pool to guarantee every client can participate in at least once and at most a certain number (\(x^*\)) of rounds. Also, we want each subset to have a size in a fixed range, and the overall data distribution as uniform as possible. These requirements can be summarized as follows: 
\begin{enumerate}
    \item[i)] Each subset has total data distributed as uniform as possible; 
    \item[ii)] Subset size is in a certain range;
    \item[iii)] Each client participates at least 1, at most \(x^*\) rounds.
\end{enumerate}
We formulate subset generation as an optimization problem as follows. 
\begin{mini!}|s|
{\mathbb{S}_1,...,\mathbb{S}_T}{max\{nid(\mathbb{S}_t) | t=1,...,T\} \label{obj3}}
{\label{P3}}{}
\addConstraint{|\mathbb{S}_t| \in [n-\delta,n+\delta], t=1,...,T \label{const31}}
\addConstraint{1 \leq \sum_{t=1}^{T}x_{kt} \leq x^*, \forall k \in \mathbb{S} \label{const32}}
\end{mini!} 
In this formulation, we address the above requirement i) by minimizing the maximum non-iid degree among all the subsets. The non-iid degree of a subset \(\mathbb{S}_t\) is denoted by \( nid(\mathbb{S}_t) \) which is defined as the non-iid degree of the overall data distribution when data of clients in \(\mathbb{S}_t\) are put together, that is, the \(Nid\) function (\ref{eq:nid}) applied to the resultant vector from addition of all histograms of clients in the subset, \(n\) and \(\delta\) are a desired number of clients and its tolerance of each subset, and
\begin{equation}
    x_{kt} = 
    \begin{cases}
      1 & \text{if \(k\in\mathbb{S}_t\)}\\
      0 & \text{else}
    \end{cases} 
\end{equation}

The constraints (\ref{const31}), (\ref{const32}) provide relaxation on the subset size and number of times a client can be selected.



\section{Proposed Solutions}\label{sec:solution}
In this section, we describe our proposed algorithms to solve the formulated problem in each stage. 
\subsection{A Greedy Algorithm to Select the Initial Client Pool}
To simplify this problem, the constraint (\ref{const13}) can be first taken care of by filtering scores based on the minimum requirements. Let \(\mathbb{K}_f\subseteq{\mathbb{K}}\) denote the set of filtered clients. Then, the constraint (\ref{const12}) can satisfied by selecting the budget \(B\) to be greater than or equal to the sum of top \(n^*\) cost values among clients in \(\mathbb{K}_f\).

\begin{equation}
    B \geq \sum_{rank i = 1}^{n^*} Cost_{rank i}
\end{equation}

With this budget, the problem becomes a 0-1 knapsack problem as follows. 

\begin{maxi!}|s|
{\mathbb{S}\subseteq{\mathbb{K}_f}}{\sum_{k\in \mathbb{S}}{ Score_k}\label{obj2}}
{\label{P1.5}}{}
\addConstraint{\sum_{k\in \mathbb{S}}{Cost_k} \leq B \label{const1.51}}
\end{maxi!} 

The well-known dynamic programming algorithm \cite{dasgupta2008algorithms} gives an exact solution with a time complexity of \(O(nB)\) where \(n = \mathbb{K}_f\). A more efficient greedy algorithm based on decreasing ratio of score to cost runs in \(O(nlogn)\) and gives an approximation within \(n^{-1/2}\) \cite{calvin2003average}. In this article, we use the greedy algorithm which works as follows: select as many clients as possible in non-increasing order of score-cost ratio. We provide an experiment in Section \ref{sec:evaluation} to show the performance of the greedy algorithm.

\subsection{Subset Generation Algorithm}


To solve the optimization problem of subset generation (\ref{P3}),  an algorithm is proposed where we select from the input pool one subset at a time, making sure that the non-iid degree of the subset is minimized (\ref{obj3}), the subset size satisfies (\ref{const31}), and keeping track of the number of times each client is selected to satisfy (\ref{const32}). The next subset is selected from the clients who have not been selected before, and the selection terminates when all the clients have been selected once or more. 

\textbf{Multidimensional Knapsack Problem (MKP)}
The selection of one subset from a given pool is formulated into a 0-1 multidimensional knapsack problem (MKP) \cite{chu1998genetic, varnamkhasti2012overview, cacchiani2022knapsack} with subset size constraints. Specifically, 
a client is treated as an item, and its data histogram $h$, specifying how many samples of each label the client has, is treated as c-dimensional weight where c is the number of classes for the classification task, each dimension (class/label) corresponding to a knapsack. 
The objective is to maximize the total data sample size of all selected clients, where the same capacity is set for all the knapsacks to encourage even distribution of data samples (weights) over classes (knapsacks). In particular, the MKP is formulated as follows. 


\begin{maxi!}|s|
{}{\sum_{k \in \mathbb{S}} |\vectorbold*{h_{k}}|_1 x_k \label{obj4}}
{\label{P4}}{}
\addConstraint{A\vectorbold*{x} \leq \vectorbold*{b} \label{const41}}
\end{maxi!} 
where \( |\vectorbold*{h_{k}}|_1 = \sum_{j=1}^{c} h_{k,j} \) is the size of the data at client 
\(k\) and simply denoted as \( |\vectorbold*{h_{k}}| \) hereinafter, \(x_k\) is a binary variable denoting whether client \(k\) is selected or not, \(\vectorbold*{x}\) is a vector of \(x_k\)s for all clients, \(\vectorbold*{b}\) is a vector of capacities of knapsacks, and \(A\) is the constraint matrix for the MKP problem which is constructed by putting together histograms of all clients, that is, $A = [h_1, h_2, ..., h_K]$ 
where \(h_k = (h_{k,1}, h_{k,2}, ..., h_{k,c})^T\) is the histogram for client \(k\), \(K = |\mathbb{S}|\) is the number of clients, and \(c\) is the number of classes or knapsacks. 

The subset size constraints are applied by adding a row of 1s and a row of -1s to \(A\) and correspondingly adding max and min sizes (\(n \pm \delta\)) to the capacity vector \(\vectorbold*{b}\). In this way, solving this MKP problem is equivalent to finding a group of clients such that their accumulated sample sizes for respective class labels fill the knapsacks as evenly as possible, thus minimizing non-iid degree of the selected group. MKP is a well-known NP-hard problem and there are several open-source solvers available, such as PuLP, Gurobi, and IBM CPLEX. We use IBM CPLEX to solve our instances of MKP. The main process of our subset generation algorithm is to select the first subset of clients from the input client pool by solving a MKP, and the second subset from the rest of the clients by solving another MKP, and so on untill no clients are left. 

\textbf{Nid Improvement:}
As subsets are selected one by one, remaining clients become fewer, so the next subset is selected from a smaller pool. This will lead to less optimal solutions to later MKPs, which means larger non-iid degrees of subsets. To mitigate this issue, we introduce $Nid$ improvement process in which $Nid$ of a selected subset is improved by adding some clients, who have been selected previously but still available, to fill in knapsacks as needed. This is made possible by utilizing constraint (\ref{const32}) which allows clients to be selected more than once. 
In particular, after selecting a subset from the current pool of remaining clients (clients who have not been selected so far), if $Nid$ of the subset is greater than a threshold, the algorithm executes the $Nid$ improvement process that: 1) finds which knapsacks are less filled (e.g., less than a certain percentage of the capacity); 2) finds clients for compensation, i.e., clients who are available for additional selection and who has data to fill the less-filled knapsacks; 3) adds these clients to the pool and selects again. 

\begin{figure}[!t]
    \centering
    \includegraphics[width=0.5\textwidth]{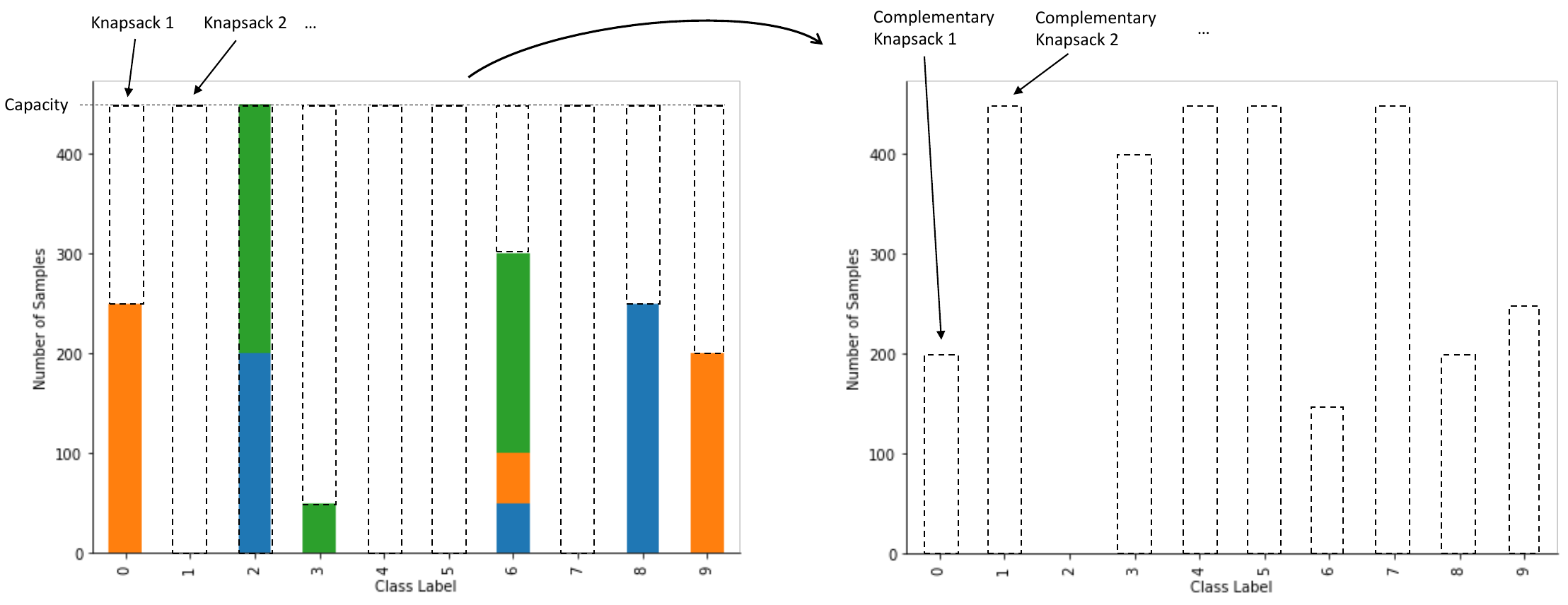}
    \caption{Complementary Knapsacks}
    \label{fig:compknap}
\end{figure}

\textbf{Complementary Knapsacks:}
To ensure that the MKP always has a solution, we relax the subset size constraint (\ref{const31}) such that the minimum size of the subset is 1 instead of $n-\delta$, and if the solution includes less than $n-\delta$ clients even after adding compensation clients, we enforce selection of at least $n-\delta$ clients and further improve $Nid$ by introducing complementary knapsacks method illustrated in Fig. (\ref{fig:compknap}). 

Suppose we want to select a fixed group of mandatory clients (items) in a MKP problem. That is, as shown on the left of Fig. (\ref{fig:compknap}), knapsacks are not all empty, some are filled to certain degrees from the beginning. Thus, the goal becomes to find other clients to fill in available spaces of the knapsacks. The trick is to formulate another MKP by setting different capacities for new knapsacks to correspond to the available spaces of current knapsacks, that is, to complement the already filled spaces, as shown on the right of Fig. (\ref{fig:compknap}), and changing the input client pool to exclude the mandatory clients. 
Complementary knapsacks are also used when too few clients are left, in which case we select all of them and find additional clients by complementary knapsacks. 

Combining the MKP formulation, $Nid$ improvement by compensation clients and complementary knapsacks, we construct our algorithm \verb|Generate Subsets| that successfully addresses subset generation problem (\ref{P3}). More detailed steps can be found in Algorithm (\ref{alg:mksubs}). 

\begin{algorithm}
\caption{Generate Subsets}\label{alg:mksubs}
\begin{algorithmic}[1]
\Require {Client pool $\mathbb{S}$; Histogram $\vectorbold*{h_k}$ of client $k \in \mathbb{S}$; Subset size $n$ and tolerance $\delta$; Max selection times per client $x^*$}
\Ensure {Subsets $\mathbb{S}_1,\mathbb{S}_2,... \subseteq{\mathbb{S}}$}

\State {Set knapsack capacities.}
\State {Define data structure to track client selection status: how many times each client has been selected.}
\State{$Remaining Clients = All Clients$}

\While{$Remaining Clients$ is not empty}
\If{$Size(Remaining Clients) \geq n-\delta$}
    \State{$Subset = SolveMKP(Remaining Clients)$.}
    \If{$Nid(Subset) > Nid Threshold$}
        \State{Improve $Nid$ by adding compensation clients.}
    \EndIf
    \If{$Size(Subset) < n - \delta$ (selected clients are too few)}
        \State {Obtain new $Subset$ that includes $n-\delta$ clients, by enforcing mandatory client selection.}
        \State{Improve $Nid$ by complementary knapsacks.}
    \EndIf
\Else{ $Size(Remaining Clients) < n-\delta$ (too few clients left)}
    \State{Select all $Remaining Clients$ into $Subset$.}
    \State{Improve $Nid$ by complementary knapsacks.}
\EndIf

\State{Save $Subset$ as $\mathbb{S}_i$.}
\State{Update $Remaining Clients$, client selection status.}
\EndWhile

\end{algorithmic}
\end{algorithm}

\section{Analysis on Fairness Guarantee}\label{sec:fairness}

We show that our proposed solutions for initial client pool selection and per-round client scheduling, 
guarantee fairness in both parts defined previously. 

Each client satisfying the minimum requirements has a chance to be included in the initial client pool. This is guaranteed by the fact that clients filtered by the minimum requirements are all considered for the optimization problem (\ref{P1.5}). Once filtered, due to the budget limit, the chance of being selected into the pool may be different between clients. 
In a scenario where the cost is given by individual clients, clients with a high score-cost ratio are more likely to be selected. So, those clients whose scores are low can also increase their chances by claiming a relatively lower price. In general, the chance relies on the client's score which is an overall score based on multiple criteria considered altogether. Thus, even if a client has a low score on some of the criteria, they can still improve the overall score by improving their scores on other criteria, thus increasing their chance of being selected. 

Once selected into the pool, each client has a similar chance to participate in each round. Our subset generation algorithm solves the MKP problem (\ref{P3}) in which each client is selected into at least one subset. So, it is guaranteed that each client participates in at least one round of each scheduling period. Further, during the \(Nid\) improvement process, a small portion of the clients may be selected into additional subsets, resulting in more than one round of participation. The proportion of these clients can be kept small by controlling the values of \(\delta\) and \(x^*\). Thus, assuming an insignificant drop-out rate, most of the clients will participate in one round per scheduling period. 


\section{Experimental Evaluation}\label{sec:evaluation}
\subsection{Experiment Settings}\label{subsec:expsetting}
For Stage 1 \textit{Initial Client Pool Selection for FL Task}, we created virtual clients by randomly assigning them scores defined in Section (\ref{sec:metric}) and compared performance of dynamic programming (DP) algorithm, random selection algorithm and the proposed greedy algorithm for selecting initial client pool from the created clients based on their scores. 

For Stage 2 \textit{Per-round Client Scheduling} We trained CNN models using MNIST and CIFAR-10 datasets distributed across 100 clients in a non-iid manner. We tested our subset generation algorithm on three different types of non-iid settings: 
\begin{itemize}
    \item Type 1 non-iid setting (one label), each client has data samples for one class label;
    \item Type 2 non-iid setting (two labels 9:1), each client has data samples distributed over two different labels with ratio 9:1;
    \item Type 3 non-iid setting (three labels 5:4:1), most clients have data samples distributed over three different labels with ratio 5:4:1, and a few clients have data samples distributed over two different labels with ratio 5:1 or 4:1. 
\end{itemize}
In each type of non-iid setting, we selected $10 \pm 3$ clients as a subset to participate in a round, resulting in 10 to 20 subsets per scheduling period during which every client gets to participate at least one round.


The experiments were conducted in a Python environment with common ML libraries such as PyTorch, TensorFlow, Keras, etc. The basic federated learning process was implemented using a public code repository \cite{jadhav2019federated} on top of which we integrated our Algorithm (\ref{alg:mksubs}) \verb|Generate Subsets| to schedule clients for each round. Further details and results are presented in the subsections below. 

\subsection{Stage 1: Initial Client Pool Selection for an FL Task}
In Experiment 1, to evaluate the performance of the score-cost ratio-based greedy algorithm, we generated 10 clients with random scores, calculated costs using the formula \ref{cost}, with \(a=2, b=5\), that is, \(Cost = 2Score + 5\) rounded to the nearest integer for convenience, and applied the dynamic programming (DP), the greedy algorithm, and random selection under the same cost budget \(B=100\). The random selection algorithm randomly selects clients until the budget is short.

Table \ref{table2} and Table \ref{table3} show the input and results of Experiment 1. The dynamic programming (DP) algorithm gives the optimal solution of 6 clients with a total score of 36.85, the greedy algorithm selects 5 clients with a total score of 32.78 and an approximation ratio of 0.11 compared to the optimal solution, and the random selection results in 28.26 total score and 0.23 approximation ratio. From this example, it can be seen that, in terms of maximizing the total score, DP is undoubtedly the best, and the greedy algorithm outperforms random selection by achieving about 90\% of the best total score. In the next experiment, we compare the time efficiencies of these three algorithms.

\begin{table}[!t]
    \centering
    \scriptsize
    \caption{Experiment 1 Input}
    \label{table2}
    \scalebox{0.92}{
    \begin{tabular}{|c|c|c|c|c|c|c|c|c|c|c|} 
     \hline
     \textbf{Client} & 0 & 1 & 2 & 3 & 4 & 5 & 6 & 7 & 8 & 9\\ 
     \hline
     \textbf{Score} & 6.92 & 4.89 & 6.8 &  6.08 & 6.9 &  6.08 & 3.74 & 3.36 & 5.26 & 3.39 \\ 
     \hline
     \textbf{Cost} & 18 & 14 & 18 & 17 & 18 & 17 & 12 & 11 & 15 & 11 \\ 
     \hline
    \end{tabular}  
    }
\end{table}

\begin{table}[!t]
    \centering
    \scriptsize
    \caption{Experiment 1 Results}
    \label{table3}
    \scalebox{1.05}{
    \begin{tabular}{|c|c|c|c|}
        \hline
         & Selected Clients & Total Score & Approx. Ratio  \\
        \hline
        Dynamic Programming & 8, 5, 4, 2, 1, 0 & 36.85 & 0\\
        \hline
        Greedy Algorithm & 0, 4, 2, 5, 3 & 32.78 & 0.11\\
        \hline
        Random Selection & 2, 1, 5, 7, 6, 9 & 28.26 & 0.23\\
        \hline
    \end{tabular}
    }
\end{table}

\begin{figure}[!t]
    \centering
    \subfigure[Our v.s. DP]{
        \includegraphics[width=.225\textwidth]{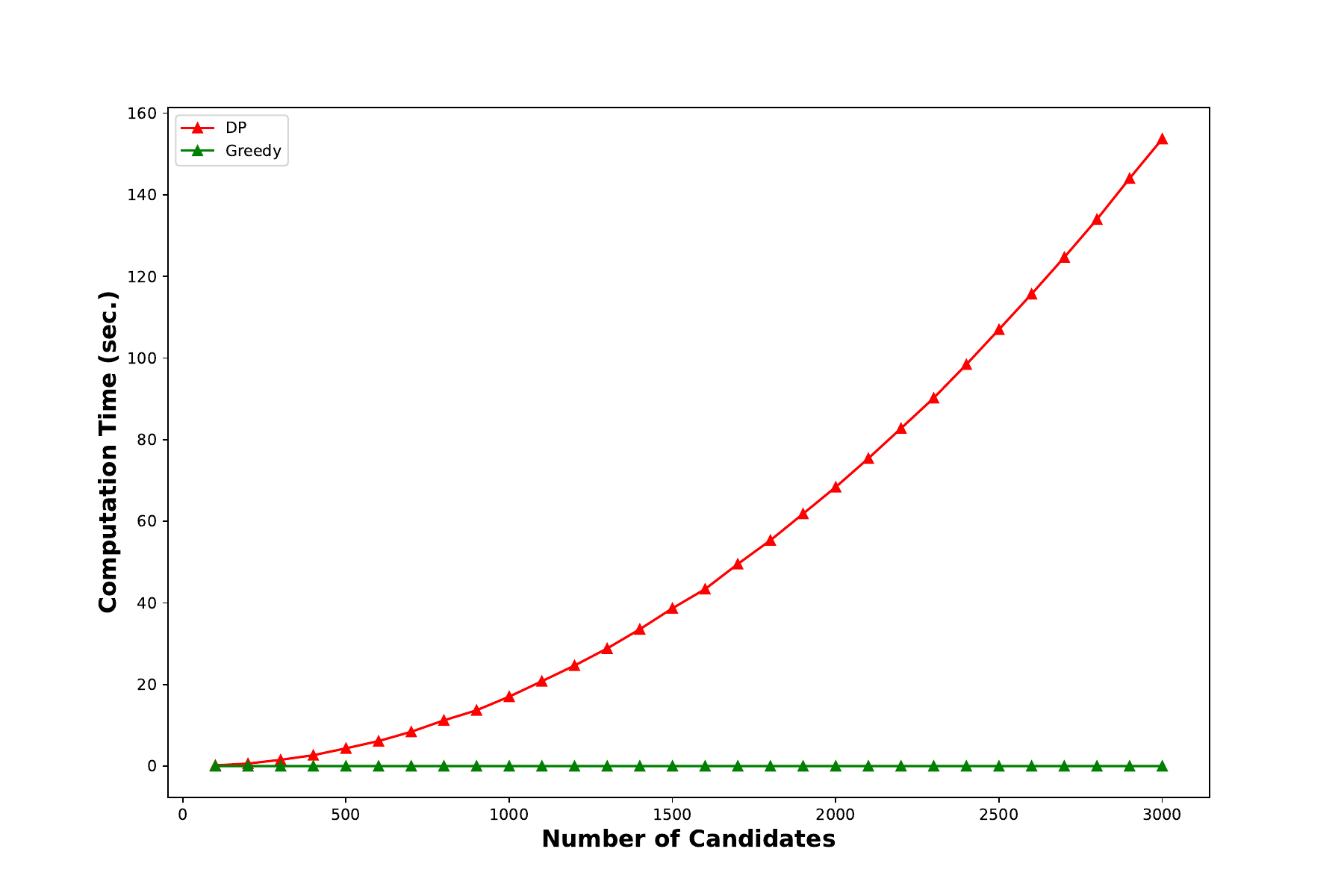}
        \label{fig:time1}}
    \hfill
    \subfigure[Our v.s. Random]{
        \includegraphics[width=.225\textwidth]
        {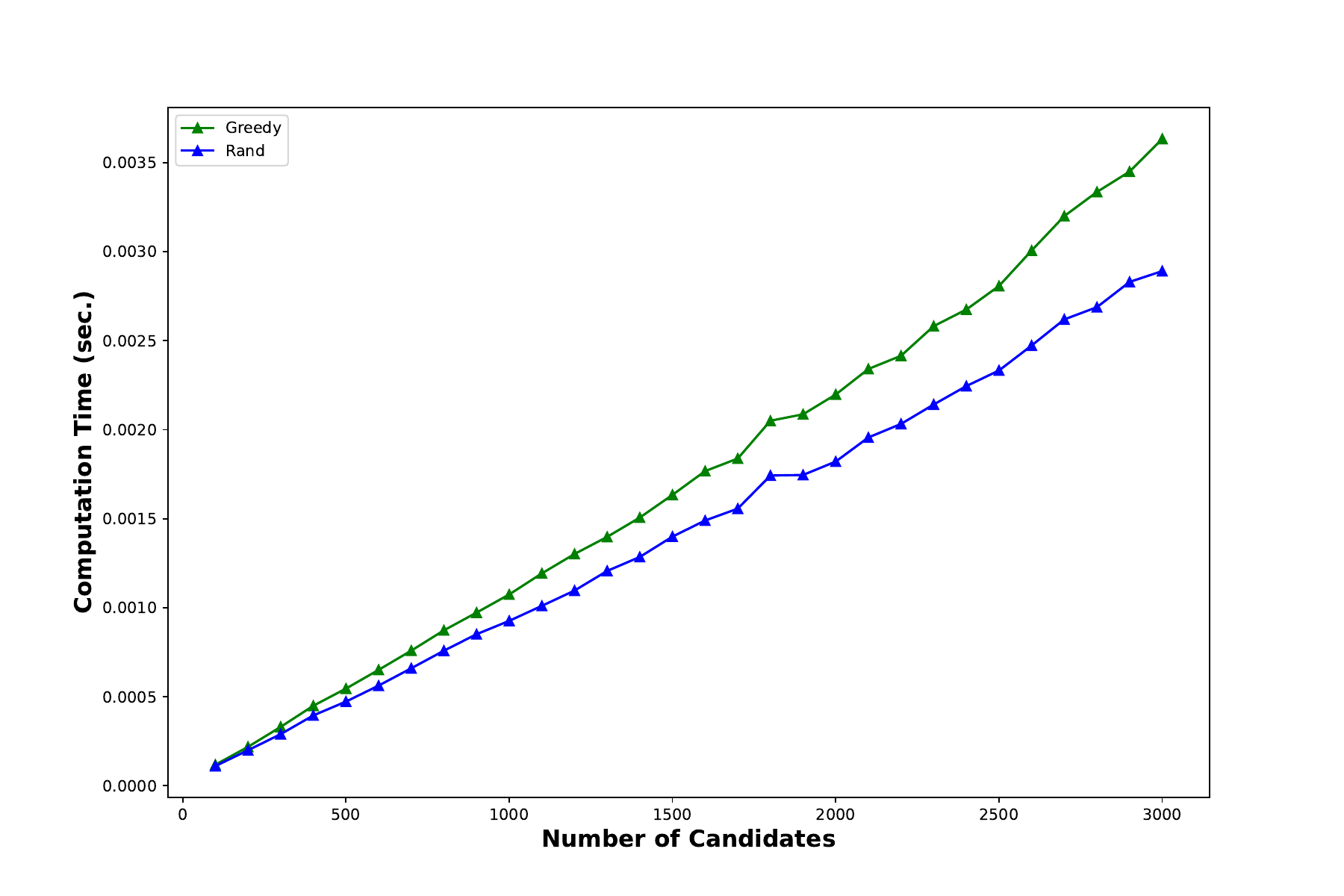}
        \label{fig:time2}}
    \caption{Computation Time vs. \# of Candidate Clients}
\end{figure}

Experiment 2 measures computation time of DP, greedy, and random selection algorithms for different number of candidate clients. The scores and costs of the candidate clients are set in the same manner as in Experiment 1. The cost budget is set proportional to the number of clients (candidates). Figs. \ref{fig:time1} and \ref{fig:time2} show how computation time changes with the number of candidates. Fig. \ref{fig:time1} plots the computation time of the dynamic programming (DP) and greedy algorithm (Greedy). As expected, DP time observes quadratic  increase with the number of candidates, while the greedy algorithm runs in almost the same time, compared to DP, across different sizes of input clients. Fig. \ref{fig:time2} shows how the greedy algorithm looks compared with the random selection. Both algorithms are very fast but the greedy algorithm becomes slightly slower than the random as the number of clients increases, which aligns with our expectation since the greedy algorithm has a time complexity of \(O(nlogn)\) and the random has that of \(O(n)\) (\(n\) being the number of candidates). 




\subsection{Stage 2: Per-round Client Scheduling}
We tested the subset generation algorithm on input client pools generated according to the three types of non-iid settings defined previouly. For each non-iid type, an input pool consisting of 100 clients is generated by assigning a data distribution histogram for each client. 

Each of the three types of client pools is fed into Algorithm (\ref{alg:mksubs}) to obtain a group of subsets for one scheduling period. Since the algorithm optimizes non-iid degree for each subset by re-selecting some clients, the resulting number of subsets may not be exactly $T=|\mathbb{S}|/n$, but greater than $T$, mostly between $T$ and $2T$. In our experiments where the size of the pool $|\mathbb{S}|=100$, subset size $n \pm \delta = 10 \pm 3$ and max selection times $x^* = 3$, the algorithm generated 10-20 subsets most of the time. 

The first step of Algorithm (\ref{alg:mksubs}) is setting capacities of knapsacks.We set one capacity for all the knapsacks since we want them evenly filled. 
The capacity is set
according to the client pool and desired number of rounds $T$. To distribute clients' data evenly across subsets, we set the capacity such that the knapsacks can accommodate data from the maximum class (label) which is the class of which data is the most abundant across the client pool. 

\begin{figure*}[!t]
    \centering
    \includegraphics[width=1\textwidth]{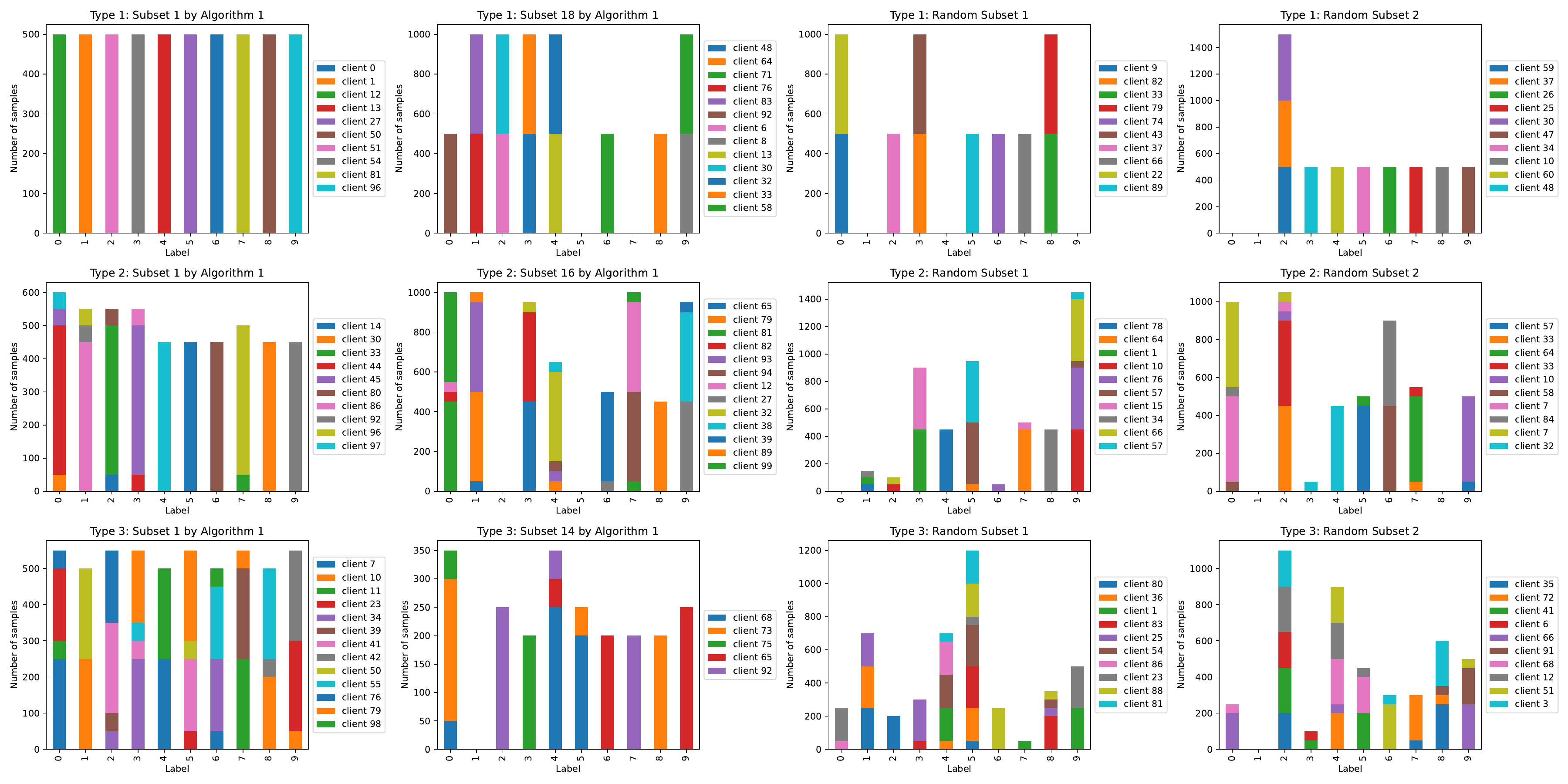}
    \caption{Subsets Generated by Algorithm \ref{alg:mksubs} and Random Subsets from Types 1-3 Pools}
    \label{fig:subsets_all}
\end{figure*}

Fig. \ref{fig:subsets_all} shows examples of subsets generated by Algorithm \ref{alg:mksubs} (left half) and subsets randomly selected (right half) from Types 1-3 input pools. Each bar graph indicates the total number of samples for each class label, where different colors represent data from different clients. For each pool type, the first and last subsets from Algorithm \ref{alg:mksubs} are shown, the rest subsets are similar to the first. Most Algorithm \ref{alg:mksubs} subsets have a close-to-uniform data distribution over labels, except for the last one which is due to the lack of remaining clients towards the end of the process. Random subsets are obtained by randomly selecting 10 clients from the pool, which have stacked histogram far away from uniform distribution. 

Based on our subset generation algorithm, we scheduled clients for each round to train a FL model according to 4 steps described in Section \ref{stage2_def}, one subset participating one round respectively, and then, the client pool is updated and a new set of subsets are generated for the next scheduling period. The pool update is simplified such that randomly selected 5\% of the clients drop out from the pool and come back after one scheduling period. For comparison, we also trained the same model using random client selection. 

\begin{figure}[]
    \centering
    \includegraphics[width=0.45\textwidth]{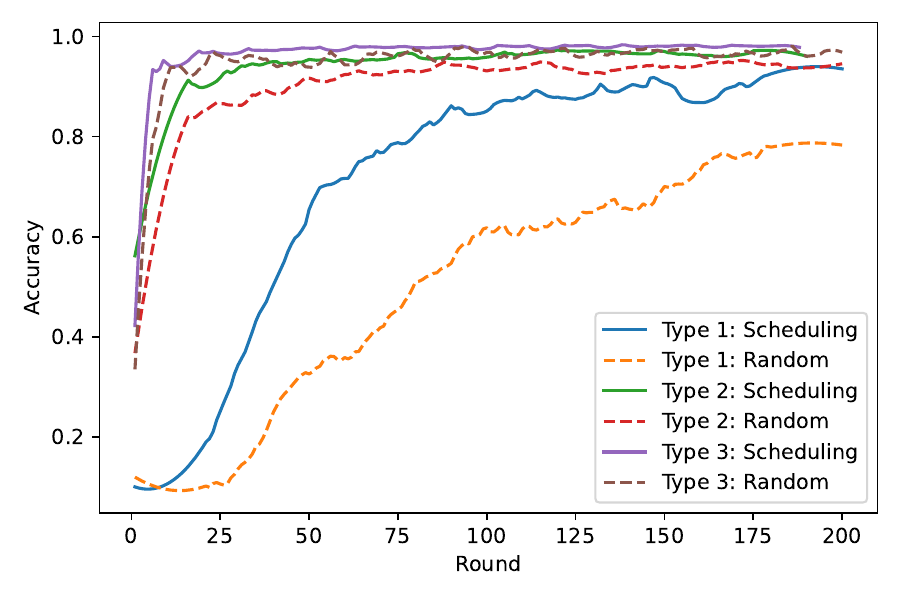}
    \caption{MNIST Learning Curves Trained from Type 1-3 Pools}
    \label{fig:mnist_curv}
\end{figure}

\begin{figure}[]
    \centering
    \includegraphics[width=0.45\textwidth]{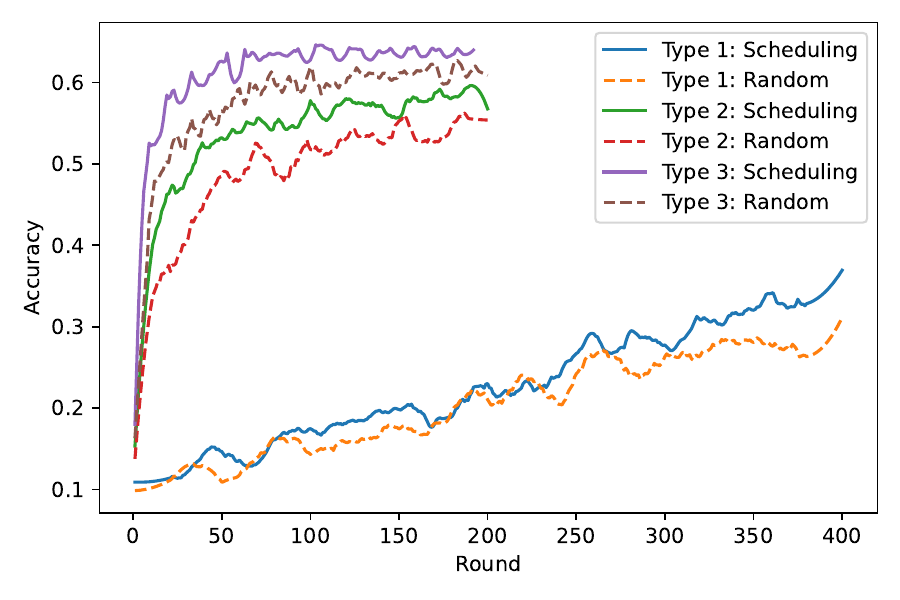}
    \caption{CIFAR Learning Curves Trained from Type 1-3 Pools}
    \label{fig:cifar_curv}
\end{figure}

For each of the 3 non-iid types, we trained a CNN model with MNIST and CIFAR datasets based on our client scheduling and random client selection. The resulting learning curves are plotted in Figs. \ref{fig:mnist_curv} and \ref{fig:cifar_curv}. 

For MNIST dataset (Fig. \ref{fig:mnist_curv}), the advantage of our scheduling method is the most obvious in Type 1 pool (each client has data of only one label), the accuracy for scheduling (green solid line) is well above that of random selection (blue dashed line), reaching an accuracy of 0.94 after 200 rounds compared to 0.78 of random. With Type 2 pool (two labels each client), the difference between scheduling and random selection reduced compared to Type 1, but still the accuracy converges faster and reaches a higher value (0.96 compared to 0.94) than random. In Type 3 pool (3 labels each client), the dashed almost caught up with the solid, but the solid shoots up a bit faster at early rounds, reaches a slightly higher accuracy (0.98 compared to 0.97), and stays more stable with less fluctuations.

For CIFAR dataset (Fig. \ref{fig:cifar_curv}), with Type 1 pool, we could not reach a convergence after training for 400 rounds, but with our scheduling method, the accuracy is higher and grows faster especially towards later rounds. With Type 2 and 3 pools, our method has a better curve as expected, reaching convergence faster and achieving higher accuracy across the rounds, the difference being more significant for Type 2 than for Type 3. 

For both datasets, we can observe a trend that the more non-iid data individual clients have (fewer labels), the more improvement achieved by our scheduling method compared to random selection. This indicates our scheduling method works better for more extreme non-iid settings. 

\section{Conclusion}\label{sec:conclusion}
In this paper, we focused on how to select clients fairly and improve learning performance for FL services and proposed a multi-criteria client selection and scheduling scheme with a fairness guarantee. It consists of three stages: 1) initial client pool selection, and 2) per-round client scheduling. Specifically, we first defined a client selection metric based on multiple criteria, including client resources, data quality, and client behaviors. Then, we formulated the initial client pool selection problem into an optimization problem aiming to maximize the overall scores of the initial client pool within a given budget and proposed a greedy algorithm to solve this problem. To guarantee fairness, we further formulated the per-round client scheduling problem and proposed a heuristic algorithm to generate several subsets from the client pool, while guaranteeing that the `integrated' dataset in a subset is close to a uniform distribution and every client is selected at least once to guarantee fairness. Our experimental results show that our scheme improves the model quality. In our future work, we will adjust the final payment according to the performance of the workers. 
\bibliographystyle{IEEEtranS}

\bibliography{Reference.bib}
    
\end{document}